# Superconductivity at 5 K in quasi-one-dimensional Cr-based KCr$_3$As$_3$ single crystals

Qing-Ge Mu[†,‡], Bin-Bin Ruan[†,‡], Bo-Jin Pan[†,‡], Tong Liu[†,‡], Jia Yu[†,‡], Kang Zhao[†,‡], Gen-Fu Chen[†,‡,§], and Zhi-An Ren[*,†,‡,§]

[†]Institute of Physics and Beijing National Laboratory for Condensed Matter Physics, Chinese Academy of Sciences, Beijing 100190, China
[‡]School of Physical Sciences, University of Chinese Academy of Sciences, Beijing 100049, China
[§]Collaborative Innovation Center of Quantum Matter, Beijing 100190, China

ABSTRACT: Recently a new family of Cr-based A$_2$Cr$_3$As$_3$ (A = K, Rb, Cs) superconductors were reported, which own a rare quasi-one-dimensional (Q1D) crystal structure with infinite (Cr$_3$As$_3$)$^{2-}$ chains and exhibit intriguing superconducting characteristics possibly derived from spin-triplet electron pairing. The crystal structure of A$_2$Cr$_3$As$_3$ is actually a slight variation of the hexagonal TlFe$_3$Te$_3$ prototype although they have different lattice symmetry. Here we report superconductivity in a 133-type KCr$_3$As$_3$ compound that belongs to the latter structure. The single crystals of KCr$_3$As$_3$ were prepared by the deintercalation of K ions from K$_2$Cr$_3$As$_3$ crystals which were grown from a high-temperature solution growth method, and it owns a centrosymmetric lattice in contrast to the non-centrosymmetric K$_2$Cr$_3$As$_3$. After annealing at a moderate temperature, the KCr$_3$As$_3$ crystals show bulk superconductivity at 5 K revealed by electrical resistivity, magnetic susceptibility and heat capacity measurements. The discovery of this KCr$_3$As$_3$ superconductor provides a different structural instance to study the exotic superconductivity in these Q1D Cr-based superconductors.

The recently discovered Cr-based superconductors $A_2Cr_3As_3$ (A = K, Rb, Cs) have attracted much interest besides the enthusiasm on Fe-based high-$T_c$ superconductors.[1-6] This is partly because there have been very few Cr-containing compounds exhibiting superconductivity for a century. Except for several binary Cr alloys,[7-11] the only superconductors are the lately reported ternary boride $Cr_2Re_3B$ which has a non-centrosymmetric β-Mn-type crystal structure and a $T_c$ of 4.8 K,[12] and the binary CrAs which exhibits superconductivity at 2 K by suppressing the antiferromagnetic order via the application of external high pressures above 8 kbar.[13-14] Furthermore, these 233-type $A_2Cr_3As_3$ compounds crystalize in a very particular quasi-one-dimensional (Q1D) hexagonal crystal lattice with a space group of *P*-6*m*2 (No. 187), which can be regarded as infinite Q1D $(Cr_3As_3)^{2-}$ linear chains separated by alkali-metal cations that act as charge reservoir.[1-3, 15] When replacing the $K^+$ ions with larger $Rb^+$ or $Cs^+$ ions, the superconducting $T_c$ decreases dramatically from 6.1 K to 4.8 K and 2.2 K respectively,[1-3, 15-16] and the $T_c$ also monotonically decreases under external pressures in $K_2Cr_3As_3$.[16-17] Theoretical calculations on $K_2Cr_3As_3$ predict complex multi-band electronic structure with a three-dimensional Fermi-surface pocket in addition to two Q1D Fermi-surface sheets mainly contributed by the Cr-3d electrons.[18-22] Experimental results show strong electron correlations and magnetic fluctuations but very diverse pictures of ground states, and both spin-triplet and spin-singlet electron pairing are proposed to explain the exotic superconductivity.[23-30] As a rare case of Q1D superconductors, the $A_2Cr_3As_3$ is significantly different from previously reported Q1D superconductors such as $Li_{0.9}Mo_6O_{17}$,[31-32] $Tl_2Mo_6Se_6$[33] and organic superconductors $(TMTSF)_2X$ (TMTSF = tetramethyltetraselenafulvalene, X = $PF_6$ or $ClO_4$).[34-37]

By deintercalating half of the K ions from $K_2Cr_3As_3$ lattice, a new type of Q1D compound $KCr_3As_3$ can be obtained.[38] This 133-type $KCr_3As_3$ has a hexagonal $TlFe_3Te_3$-type crystal structure with the space group *P*6$_3$/*m* (No. 176).[38] Unlike its 233-type cousins which lack of inversion symmetry, the $KCr_3As_3$ has a centrosymmetric crystal lattice. Although the characteristic $Cr_3As_3$ linear chain structures are similar between the two compounds except a small-angle rotation along the *c*-axis, the removing of a $K^+$ ion per formula from $K_2Cr_3As_3$ increases the chemical valence state of Cr and turns the Q1D chains into $(Cr_3As_3)^-$ type in $KCr_3As_3$ lattice. Previous experimental studies on polycrystalline $KCr_3As_3$ showed cluster spin-glass ground state without superconductivity at low temperature,[38] and the DFT first-principle calculations exhibited a magnetic Fermi surface involving only three one-dimensional sheets with much reduced dimensionality and the emergence of the interlayer antiferromagnetic order.[39]

In this communication, we report the discovery of superconductivity in the single crystals of $KCr_3As_3$ with a $T_c$ of 5 K.

The 133-type KCr$_3$As$_3$ single crystals were prepared by the deintercalation of K$^+$ ions from K$_2$Cr$_3$As$_3$ precursors. At first, high quality single crystals of K$_2$Cr$_3$As$_3$ were grown out of KAs and CrAs mixture using a high-temperature solution growth method as previously reported.[1] Then the as-grown rod-like K$_2$Cr$_3$As$_3$ single crystals were immersed in pure dehydrated ethanol and kept for one week for the deintercalation of K$^+$ ions at room temperature. The obtained samples were washed by ethanol thoroughly and labeled as sample #A. To further improve the sample quality, the crystals of sample #A were immersed in ethanol in a Teflon liner and loaded into an autoclave. The autoclave was tightly sealed, and sintered at 353 K for 100 h. After cooled down to room temperature, the obtained samples were washed with ethanol again and annealed in an evacuated quartz tube at 373 K for 12 h. These final post-annealed KCr$_3$As$_3$ crystals were labeled as sample #B. All the experimental procedures were carried out in a glove box filled with high-purity Ar gas to avoid introducing impurities. The obtained KCr$_3$As$_3$ crystals are stable in air at room temperature, but it decomposes above a moderate temperature at 473 K. Due to this reason, our direct synthesis of KCr$_3$As$_3$ polycrystals from solid state reaction or single crystal growth from high-temperature solutions all failed. For comparison, we also reproduced polycrystalline KCr$_3$As$_3$ samples by the deintercalation process from K$_2$Cr$_3$As$_3$ powders which were synthesized by a solid state reaction method as reported.[38]

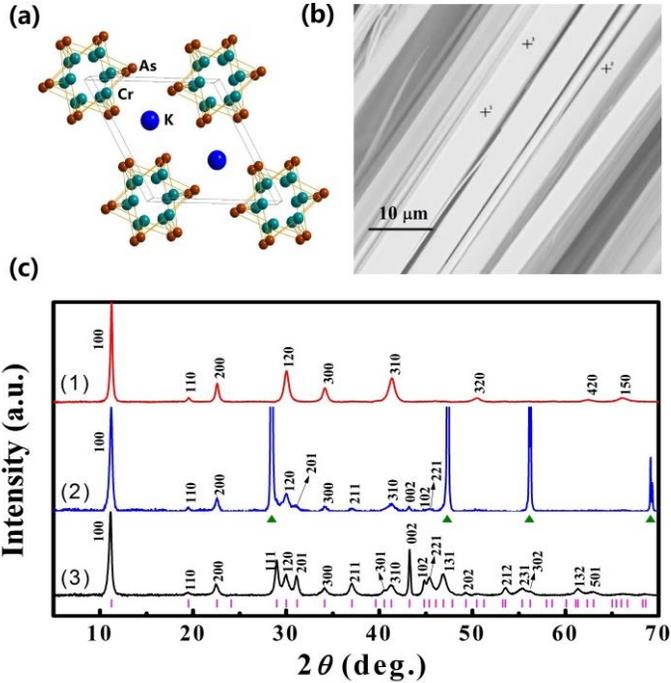

**Figure 1.** (a) The hexagonal crystal structure of KCr$_3$As$_3$. (b) The scanning electron microscope image of KCr$_3$As$_3$ crystal. (c) The XRD patterns for three KCr$_3$As$_3$ samples: (1) crystals of sample #B, (2)

ground powder of sample #B mixed with Si, (3) polycrystalline sample. The triangles represent Bragg peaks from Si, and the vertical bars represent the position of calculated Bragg peaks.

The crystal structure of all samples was characterized by powder X-ray diffraction (PXRD) at room temperature with a PAN-analytical X-ray diffractometer using Cu-K$_\alpha$ radiation. The electrical resistivity and heat capacity were measured in a Quantum Design physical property measurement system (PPMS) by the standard four-probe method and relaxation method respectively. The dc magnetization was measured in a Quantum Design magnetic property measurement system (MPMS) under zero-field-cooling (ZFC) and field-cooling (FC) modes.

The hexagonal crystal structure of the 133-type KCr$_3$As$_3$ is illustrated in Figure 1a. Due to the obvious lattice shrinkage along the $a$-axis during the ion deintercalation process (about 8.9% from the K$_2$Cr$_3$As$_3$ precursor), the crystals of KCr$_3$As$_3$ split into threadlike filaments with micrometer-size diameters and millimeter-size lengths as shown by the scanning electron microscope image in Figure 1b, differing from the solid rod-like K$_2$Cr$_3$As$_3$ crystal. This crystal morphology makes it incapable for a fully single-crystal X-ray diffraction analysis for structural determination; hence PXRD is employed for structural study. Figure 1c shows the XRD patterns for three samples, which reveal consistent diffraction peaks with identical crystal structure. The XRD patterns of polycrystalline sample indicate a pure single phase of the 133 structure, as indicated by the vertical bars of theoretical calculations for Bragg peak positions. The lattice parameters were refined to be $a$ = 9.092(1) Å, $c$ = 4.180(6) Å using the $P6_3/m$ (No. 176) space group, close to the previous results.[38] For the annealed single crystals of sample #B, the XRD patterns on the crystal side surface only shows the peaks of (hk0) since the sample is thread-like along $c$-axis. To determine the lattice parameters, the sample #B was ground together with silicon powder (which acts as both of abraser and internal standard) and the PXRD patterns show several additional diffraction peaks relevant to the $c$-axis. We note that the KCr$_3$As$_3$ crystal has good ductility and it is difficult to make fine powder, therefore the diffraction peaks still have highly preferred orientation. The refined lattice parameters for sample #B are $a$ = 9.090(8) Å and $c$ = 4.182(9) Å, which are very close to the results of polycrystalline samples. In addition, no obvious difference in XRD patterns for sample #A and sample #B was observed; and in all deintercalated samples, no diffraction peak from possible remnant K$_2$Cr$_3$As$_3$ phase was detected.

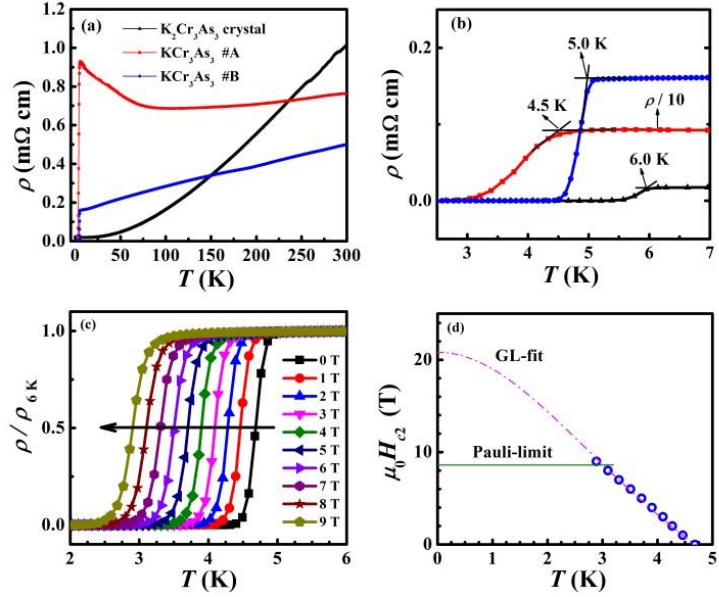

**Figure 2.** (a) Temperature dependence of electrical resistivity for crystals of $K_2Cr_3As_3$ and $KCr_3As_3$ (both sample #A and #B). (b) Enlarged view for the resistive superconducting transitions. (c) The superconducting transitions under different magnetic fields for sample #B from 0 T to 9 T. (d) Derived upper critical field and the Pauli paramagnetic limit for sample #B.

The temperature dependence of electrical resistivity was characterized from 1.8 K to 300 K for all samples and the data are shown in Figure 2. In our experiments, all batches of $KCr_3As_3$ single crystals show superconducting transitions with resistivity dropping to zero at low temperatures. But for the polycrystalline samples, no superconductivity was observed as previously reported.[38] In Figure 2a, we compared the resistivity behavior between the single crystals of $K_2Cr_3As_3$ and $KCr_3As_3$ (both sample #A & #B). The room temperature resistivity values are close for all these samples, but the residual resistance ratio (RRR) of the $KCr_3As_3$ (~ 3 for sample #B) is much smaller than that of $K_2Cr_3As_3$ (~ 60), indicating the poor crystalline quality of $KCr_3As_3$ crystals. This can be explained by the crystal defects and lattice deformation in the $KCr_3As_3$ crystal induced by the deintercalation process around room temperature, which also make the sample #A semiconducting-like behavior below 100 K. While after annealing, the crystal quality becomes better and it shows a metallic resistivity behavior for sample #B. This can be further illustrated by the superconducting transition as shown in Figure 2b. The sample #A shows a wide superconducting transition and an onset $T_c$ ~ 4.5 K, while the sample #B shows a much narrower superconducting transition with a higher onset $T_c$ ~ 5.0 K, and both values are lower than the $T_c$ ~ 6.0 K of $K_2Cr_3As_3$.

To further characterize the superconducting properties, we performed resistivity measurements on sample #B under different magnetic fields from 0 T to 9 T to study the upper critical field $H_{c2}$ (with the field perpendicular to the $c$-axis and electrical current along the $c$-axis), and the normalized data for $\rho/\rho_{6K}$ vs. $T$ are shown in Figure 2c. With magnetic field increasing, the $T_c$ shifts to lower temperatures systematically. We define the $\mu_0H_{c2}$ as the field determined by 50% of the normal-state resistivity at $T_c$, and it was depicted as a function of temperature in Figure 2d. Considering Ginzburg-Landau (GL) theory, $\mu_0H_{c2}(T) = \mu_0H_{c2}(0)(1 - t^2)/(1 + t^2)$, here $t = T/T_c$, the zero-temperature upper critical field $\mu_0H_{c2}(0)$ is estimated to be 20.8 T. This value is much higher than the Pauli paramagnetic limited upper critical field $\mu_0H_p = 1.84T_c \approx 8.6$ T,[40] which gives an evidence for unconventional superconductivity in KCr$_3$As$_3$, and similar phenomenon was also reported in Li$_{0.9}$Mo$_6$O$_{17}$,[41] Sr$_2$RuO$_4$[42] and K$_2$Cr$_3$As$_3$ superconductors.[1]

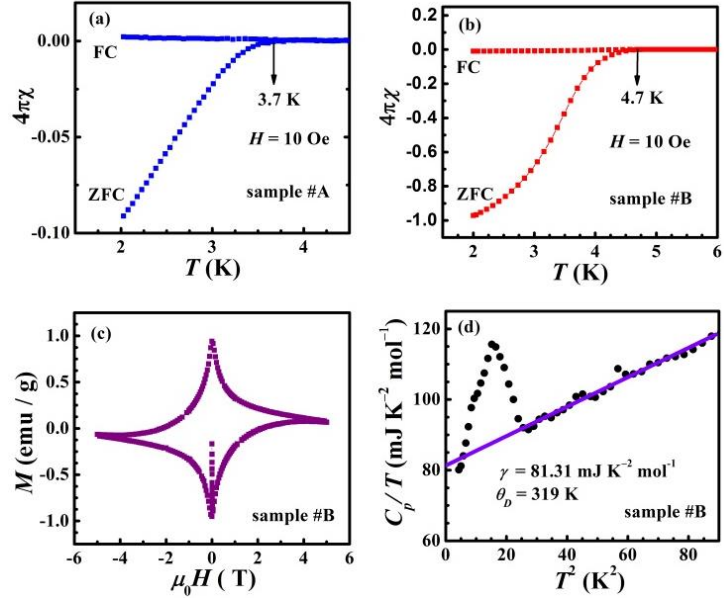

**Figure 3.** Temperature dependence of magnetic susceptibility for (a) sample #A and (b) sample #B. (c) Isothermal magnetization for sample #B at 2 K. (d) Low temperature heat capacity for sample #B depicted as $C_p/T$ vs. $T^2$ with a normal state linear fit.

To demonstrate whether the observed superconductivity is the bulk nature of the K-133 phase or from possible minor remnant K-233 phase which cannot be detected by XRD, the temperature dependence of magnetic susceptibility and heat capacity were characterized and shown in Figure 3. Under a magnetic field of 10 Oe (perpendicular to the $c$-axis), the sample #A and #B show clear diamagnetic supercon-ducting transition at 3.7 K and 4.7 K respectively. While the shielding volume fraction at 2 K from the ZFC data is only about 9% for sample #A, and it is significantly enhanced to nearly 98% for sample #B. This is consistent with the behavior of resistive superconducting transition. The results indicate that the sample #A only shows very poor superconductivity with a small superconducting fraction, while after annealing, as the crystal lattice is reformed, the sample #B becomes a good bulk superconductor. For the normal state susceptibility of sample #B, it roughly coincides with the Curie-Weiss behavior with no magnetic ordering transition. In Figure 3c we show the isothermal magnetization curve of sample #B with respect to magnetic field from -5 T to 5 T at 2 K, and it reveals typical type-II superconductivity in single crystalline $KCr_3As_3$.

The temperature dependence of heat capacity is plotted in Figure 3d as a relationship for $C_p/T$ vs. $T^2$. The normal state data are linearly fitted with both electron and phonon contributions by $C_p/T = \gamma + \beta T^2$ with $T$ above $T_c$, from which we obtain the Sommerfeld coefficient $\gamma$ as 81.31 mJ/(mol K$^2$), and Debye temperature $\theta_D$ as 319 K calculated according to $\theta_D = [(12/5)NR\pi^4/\beta]^{1/3}$. Comparing with $K_2Cr_3As_3$, the close $\gamma$ values indicate similar strong electron correlations, while the higher $\theta_D$ is reasonably originated from the condensed crystal lattice for $KCr_3As_3$.[1,16] The clear heat capacity jump in the $C_p$ curve happens at about 5 K. This indicates the occurrence of superconducting transition and further confirms the bulk superconductivity of $KCr_3As_3$ single crystals, which is consistent with the results of resistivity and magnetization measurements. The similar superconducting and electronic characteristics in $KCr_3As_3$ and $K_2Cr_3As_3$ possibly indicate same electron pairing mechanism that derives from different lattice symmetry, and the lack of superconductivity in polycrystalline $KCr_3As_3$ and the annealing effects for single crystals reveal the extreme sensitivity of superconductivity by disorders in the crystal lattice.

In summary, we synthesized the 133-type $KCr_3As_3$ single crystals, and found bulk superconductivity at a $T_c$ of 5.0 K. This compound has a centrosymmetric crystal structure differing from its non-centrosymmetric counterpart $K_2Cr_3As_3$. Considering similar superconducting characteristics but different crystal symmetry between the two superconductors, this $KCr_3As_3$ provides another platform to acquire deep insight into the unconventional superconducting mechanism in these Q1D Cr-based superconductors.

**Corresponding Author**

* renzhian@iphy.ac.cn

ACKNOWLEDGMENTS

The authors are grateful for the financial supports from the National Natural Science Foundation of China (No. 11474339), the National Basic Research Program of China (973 Program, No. 2016YFA0300301) and the Youth Innovation Promotion Association of the Chinese Academy of Sciences.